\documentclass[twocolumn,amsmath,showkeys,prb]{revtex4-1}% Physical Review B

\usepackage{dcolumn}% Align table columns on decimal point
\usepackage{bm}% bold math
\usepackage[T1]{fontenc}
\usepackage{amsmath}
\usepackage{graphicx}

\usepackage{hyperref}
\hypersetup{colorlinks=true, linkcolor=blue, citecolor=red, urlcolor=magenta, pdftitle={Issues with J-dependence in the LSDA+U method for non-collinear magnets}, pdfauthor={Eric Bousquet}}

\begin{document}
\title{Issues with $J$-dependence in the LSDA$+U$ method for non-collinear magnets}

\author{Eric Bousquet$^{1,2}$}
\author{Nicola Spaldin$^{3}$}
\affiliation{$^1$Materials Department, University of California, 
             Santa Barbara, CA 93106, USA}
\affiliation {$^2$Physique Th\'eorique des Mat\'eriaux, Universit\'e de Li\`ege,  
              B-4000 Sart Tilman, Belgium}
\affiliation {$^3$ Department of Materials, ETH Zurich, Wolfgang-Pauli-Strasse 10 CH-8093 Zurich, Switzerland}

\begin{abstract}
We re-examine the commonly used density functional theory plus Hubbard \textit{U} (DFT$+U$) method for
the case of non-collinear magnets.
While many studies neglect to explicitly include the exchange correction parameter \textit{J}, or 
consider its exact value to be unimportant, here we show that in the case of non-collinear 
magnetism calculations the \textit{J} parameter can strongly affect the magnetic ground state.
We illustrate the strong \textit{J}-dependence of magnetic canting and magnetocrystalline anisotropy by calculating trends
in the magnetic lithium orthophosphate family LiMPO$_4$ (M = Fe and Ni) and difluorite family MF$_2$ (M = Mn, Fe, Co and Ni).
Our results can be readily understood by expanding the usual DFT$+U$ equations within the spinor scheme, 
in which the \textit{J} parameter acts directly on the off-diagonal components which determine the
spin canting. 
% Interestingly, the calculated magnetocrystalline anisotropy energy is only weakly $J$-dependent but depends strongly on the $U$ parameter.
\end{abstract}


\maketitle

Density functional theory (DFT) within the local density (LDA) and generalized gradient (GGA) 
approximations is widely used to describe a large variety of materials with good accuracy. The
LDA and GGA functionals often fail, however, to correctly reproduce the properties of strongly 
correlated materials containing \textit{d} and \textit{f} electrons.
The LDA$+U$ approach -- in which a Hubbard \textit{U} repulsion 
term is added to the LDA functional for selected orbitals --
was introduced in response to this problem, and often improves drastically over the LDA or GGA.
Indeed, it provides a good description 
of the electronic properties of a range of exotic magnetic materials, such as the Mott insulator 
KCuF$_3$\cite{liechtenstein1995} and the metallic oxide LaNiO$_2$~\cite{lee2004}.

Two main LDA$+U$ schemes are in widespread use today: The Dudarev~\cite{dudarev1998} approach in which an isotropic screened on-site Coulomb interaction $U_{eff}=U-J$ is added, and the Liechtenstein~\cite{liechtenstein1995} approach in which the $U$ and exchange ($J$) parameters are treated separately. The Dudarev approach is equivalent to the Liechtenstein approach with $J=0$~\cite{baettig2005}.
Both the effect of the choice of LDA+$U$ scheme on the orbital occupation and subsequent properties
~\cite{ylvisaker2009}, as well as the dependence of the magnetic properties on the value of $U$
~\cite{savrasov2005}, 
have recently been analyzed. There has been no previous systematic study, however, of the effect of the  
$J$ parameter of the Liechtenstein approach in non-collinear magnetic materials. 
Here we show that neither the approach of not explicitly considering the $J$ parameter (as in the Dudarev implementation),
nor the assumption that its importance is borderline -- a common approximation is to use $J\simeq10\%\ U$ without careful
testing -- within the Liechtenstein implementation are justified in the case of non-collinear magnets.
We demonstrate that in the case of non-collinear antiferromagnets, the choice of $J$ can strongly 
change the amplitude of the spin canting angle (LiNiPO$_4$) or even modify the easy axis of the system 
(LiFePO$_4$ and FeF$_2$), with consequent drastic effects on the magnetic susceptibilities and 
magnetoelectric responses.

First we remind the reader how the $U$ and $J$ parameters appear in the usual collinear
spin LSDA$+U$ formalism.
The LSDA$+U$ reformulation of the LSDA Hamiltonian is usually written as:
\begin{align}
 H_{LSDA+U}=H_{LSDA}+H_U \quad ,
\end{align}
whith
\begin{align}
 H_U^\sigma=\displaystyle\sum_{m_1,m_2}P_{m_1,m_2}V^\sigma_{m_2,m_1} \quad,
\end{align}
where $P$ is the projection operator, $\sigma$ is the spin index,  and (on a given atomic site):
\begin{align}
&V^{\uparrow(\downarrow)}_{m_2,m_1}=& \nonumber\\
& \displaystyle\sum_{3,4}\left( V^{ee}_{1,3,2,4}-U\delta_{1,2}-V^{ee}_{1,3,4,2}+J\delta_{1,2}\right)n^{\uparrow(\downarrow)}_{3,4}& \nonumber \\
& +\left( V^{ee}_{1,3,2,4}-U\delta_{1,2}\right)n^{\downarrow(\uparrow)}_{3,4}+\frac{1}{2}(U-J)\delta_{1,2} &
\label{Vupup}
\end{align}
% \begin{align}
% &V^{\downarrow}_{m_2,m_1}=& \nonumber \\
% &\displaystyle\sum_{3,4}\left( V^{ee}_{1,3,2,4}-U\delta_{1,2}-V^{ee}_{1,3,4,2}+J\delta_{1,2}\right)n^{\downarrow}_{3,4}& \nonumber \\
% &+\left( V^{ee}_{1,3,2,4}-U\delta_{1,2}\right)n^{\uparrow}_{3,4}+\frac{1}{2}(U-J)\delta_{1,2} &
% \label{Vdndn}
% \end{align}
Here $V^{ee}_{1,3,2,4}=\left\langle m_1,m_3 \left| V^{ee}_{m_1,m_3,m_2,m_4} \right| m_2,m_4\right\rangle$ are the elements of the screened 
Coulomb interaction (which can be viewed as the sum of Hartree (direct) contributions $V^{ee}_{1,3,2,4}$ and Fock (exchange) contributions 
$V^{ee}_{1,3,4,2}$
and $n^{\sigma}_{i,j}$ are the $d$-orbital occupancies.

In the case of non-collinear magnetism, the formalism is extended and the density is expressed in a two-component spinor formulation:
\begin{align}
\rho=&
\begin{pmatrix}
 \rho^{\uparrow\uparrow} & \rho^{\uparrow\downarrow} \\
 \rho^{\downarrow\uparrow} & \rho^{\downarrow\downarrow}
\label{rho}
\end{pmatrix}
= 
\begin{pmatrix}
 n+m_z &  m_x-im_y\\
 m_x+im_y & n-m_z
\end{pmatrix}
\end{align}
where $n$ is the charge density and $m_\alpha$ the magnetization density along the $\alpha$ direction ($\alpha= x, y, z$).
Using the double-counting proposed by Bultmark \textit{et al.}\cite{bultmark2009}, the LSDA$+U$ potential is then also expressed in the two-component spin space as:
\begin{align}
V_{i,j}=&
\begin{pmatrix}
 V^{\uparrow\uparrow}_{i,j} & V^{\uparrow\downarrow}_{i,j} \\
 V^{\downarrow\uparrow}_{i,j} & V^{\downarrow\downarrow}_{i,j}
\end{pmatrix}
\end{align}
where $V^{\uparrow\uparrow}$ and $V^{\downarrow\downarrow}$ are equal to Eqs.\ref{Vupup} and
\begin{align}
V^{\uparrow\downarrow(\downarrow\uparrow)}_{m_2,m_1}=\displaystyle\sum_{3,4}\left(-V^{ee}_{1,3,4,2}+J\delta_{1,2}\right)n^{\uparrow\downarrow(\downarrow\uparrow)}_{3,4}
\label{Vupdn}
\end{align}
% \begin{align}
% V^{\downarrow\uparrow}_{m_2,m_1}=\displaystyle\sum_{3,4}\left(-V^{ee}_{1,3,4,2}+J\delta_{1,2}\right)n^{\downarrow\uparrow}_{3,4}
% \label{Vdnup}
% \end{align}

For collinear magnets, only $V^{\uparrow\uparrow}$ and $V^{\downarrow\downarrow}$ (Eqs.~\ref{Vupup}) are relevant 
since $n^{\uparrow\downarrow}$ and $n^{\downarrow\uparrow}$ are equal to zero, and $J$ affects the potential mainly through an effective 
$U-J$. 
However, in the case of non-collinear magnetism, the $n^{\uparrow\downarrow}$ and $n^{\downarrow\uparrow}$ and hence the
$V^{\uparrow\downarrow}$ and $V^{\downarrow\uparrow}$ (Eqs.~\ref{Vupdn}) are non-zero. Then it is clear 
from Eqs.~\ref{Vupdn} that $J$ acts explicitly on the off-diagonal potential components.

Next, we show the effect of the choice of $J$ parameter in the family of lithium orthophosphates, LiMPO$_4$ (M = Ni and Fe) and in the family of difluorites MF$_2$ (M = Mn, Co, Fe and Ni).
The orthophosphates crystallize in the orthorhombic \textit{Pnma} space group with $C$-type antiferromagnetic (AFM) order.
The difluorites crystalize in the tetragonal \textit{P4$_2$/mnm} rutile structure with AFM order.
We performed calculations within the Liechtenstein approach of the DFT$+U$ as implemented in the VASP code~\cite{vasp1,vasp2} 
\footnote{We note that LSDA$+U$ double-couting term taking into accound the magnetization density as proposed by Bultmark \textit{et al.}\cite{bultmark2009} is mandatory within non-collinear magnetism calculations. 
This is not necessarily done in the present implementation of other codes.}
with $U$ and $J$ corrections applied to the 3\textit{d} orbitals of the M cations.
In all cases we relaxed the atomic positions until the residual forces on each atom  were lower than 10 $\mu$eV/\AA\ at the experimental volume and cell shape reported in Tab.~\ref{tab:Exp-acell}, taking into account the spin-orbit interaction.
We found good convergence of the non-collinear spin ground state with a cutoff energy of 500 eV on the plane wave expansion and a k-point grid of $2\times4\times4$ for the orthophophates and $4\times4\times6$ for the difluorites.
% In spite of what previous assertions for related calculations with spin-orbit explicitly
% included, we find that the important non-collinear properties (magnetocrystalline anisotropy, 
% spin canting, magnetoelectric responses, etc.) do not need large numbers of k-points or 
% plane waves for convergence.

\begin{table}[htbp!]
\begin{center}
\begin{tabular}{lcccc}
\hline
\hline
                    & $a$     & $b$    & $c$     &  Ref.    \\
\hline
%  LiMnPO$_4$         & 10.44 & 6.09 & 4.75  &  \onlinecite{geller1960}  \\
 LiFePO$_4$         & 10.332& 6.010& 4.692 &  \onlinecite{streltsov1993}   \\
%  LiCoPO$_4$         & 10.202& 5.922& 4.699 &  \onlinecite{amine2000}  \\
 LiNiPO$_4$         & 10.032& 5.854& 4.677 &  \onlinecite{abrahams1993}  \\
NiF$_2$             & 4.650 & 4.650& 3.084 &  \onlinecite{hutchings1970} \\
FeF$_2$             & 4.700 & 4.700& 3.310 &  \onlinecite{dealmeida1989} \\
MnF$_2$             & 4.650 & 4.650& 3.084 &    \onlinecite{oguchi1958} \\
CoF$_2$             & 4.695 & 4.695& 3.179 &  \onlinecite{otoole2001}\\
\hline
\end{tabular}
\caption{Experimental cell parameters (\AA) used in the simulations of LiMPO$_4$ phosphates and MF$_2$ difluorites.}
\label{tab:Exp-acell}
\end{center}
\end{table}

\begin{figure}[ht]
 \includegraphics[width=6.3cm,keepaspectratio=true]{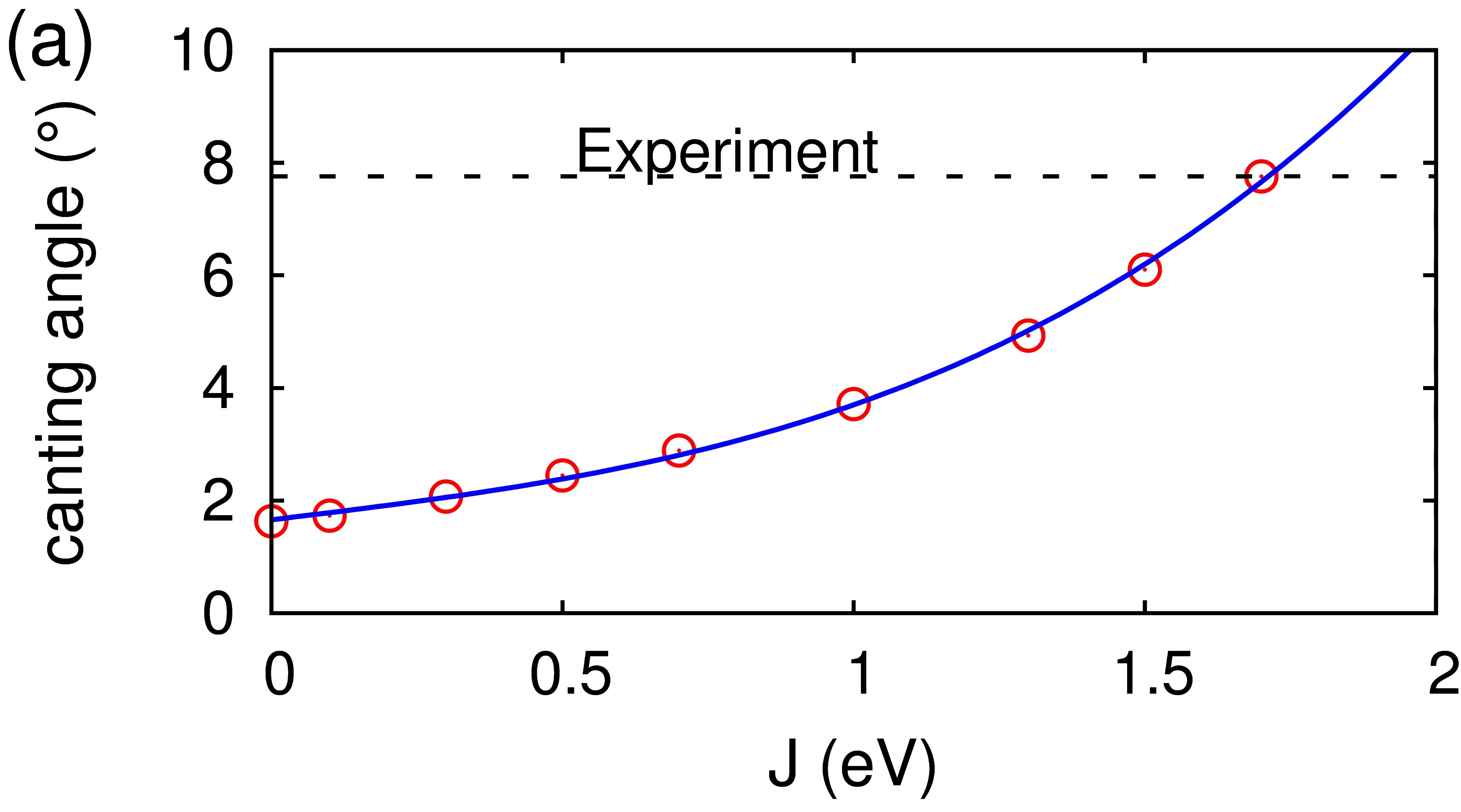}
 \includegraphics[width=6.5cm,keepaspectratio=true]{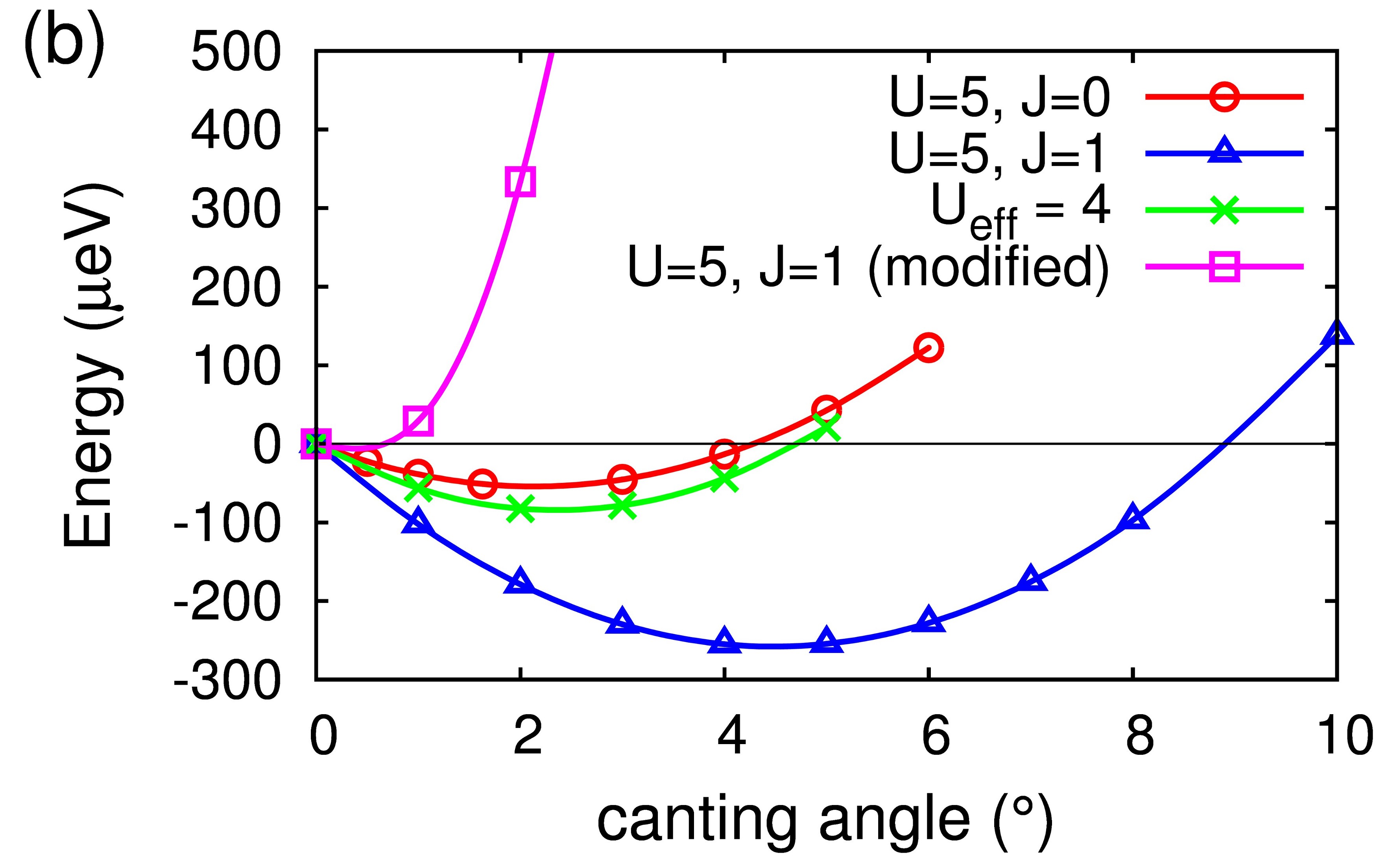}
 \includegraphics[width=6.5cm,keepaspectratio=true]{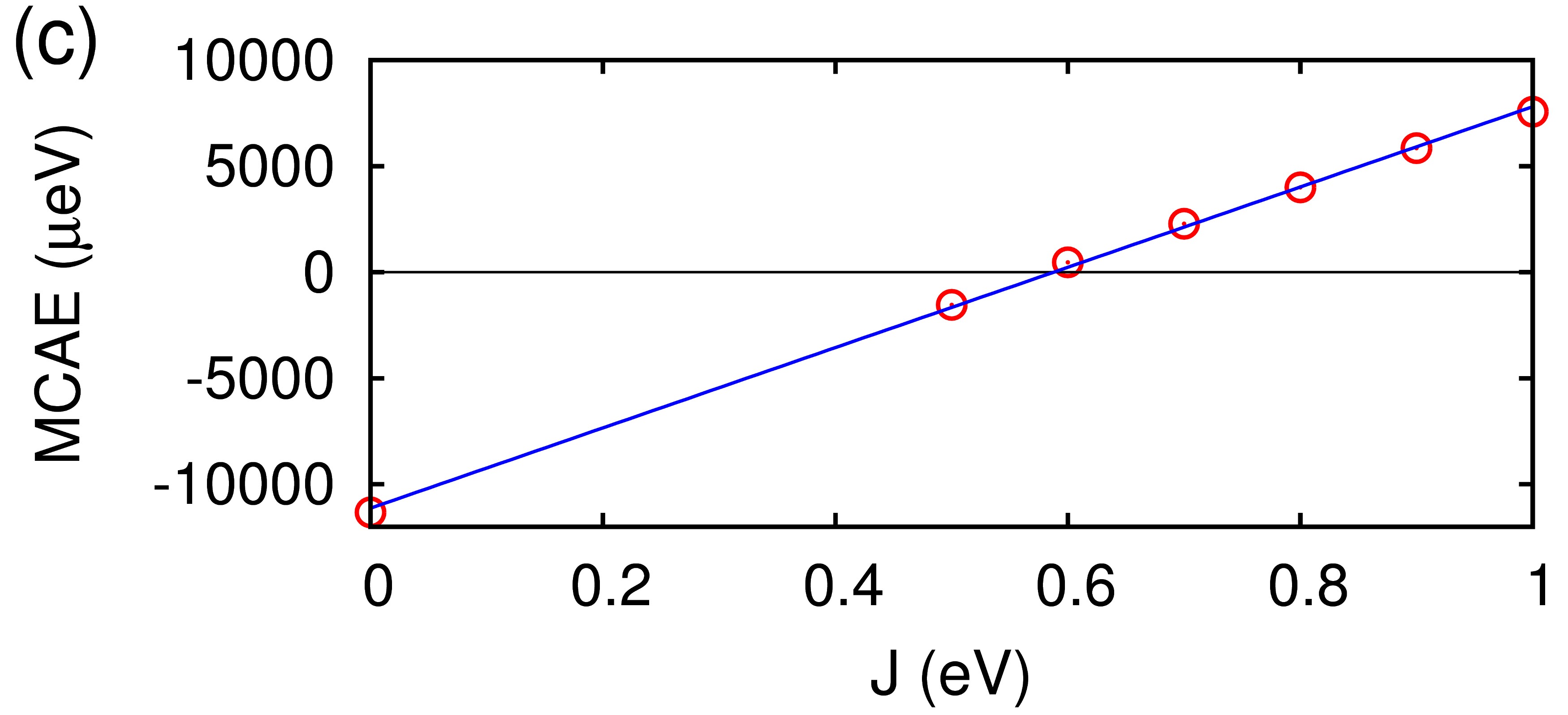}
% m_vs_J_LDA_LiNiPO4.pdf: 792x612 pixel, 72dpi, 27.94x21.59 cm, bb=0 0 792 612
 \caption{(a) Calculated LSDA$+U$ canting angle of LiNiPO$_4$ versus $J$ for $U=5$ eV. 
 The experimental value of the canting angle is equal to 7.8$^\circ$~\cite{jensen2009}.
(b) Energy versus canting angle in LiNiPO$_4$ for $U=5$ eV and $J=0$ eV (red circles), $U=5$ eV and $J=1$ eV (blue triangles), $U_{eff}=4$ eV (green crosses) and $U=5$ eV and $J=1$ eV but by fixing $J=$ 0 eV in Eqs.\ref{Vupdn} (pink squares). The zero energy reference is chosen at zero canting angle.
(c) Magnetocrystaline anisotropy energy (MCAE) between the $a$ and $b$ orientations of the magnetic moments of LiFePO$_4$.
The experimental $b$ orientations is taken as energy reference.}
 \label{fig:LiMPO4}
\end{figure}

First, we focus on LiNiPO$_4$, which is known experimentally
to be $C$-type AFM, with an easy-axis along the $c$ direction and a small A-type AFM canting of the spins along the $a$ direction ($C_zA_x$ ground state with \textit{mm'm} magnetic point group)~\cite{jensen2009}.
Performing calculations within the LSDA$+U$ method with $J=0$, we find that we correctly reproduce the $C_zA_x$ ground state 
with a rather small $U$ sensitivity of the magnetocrystalline anisotropy energy (MCAE) and the spin canting; this finding is consistent
with a previous report using the GGA functional~\cite{yamauchi2010}.
However, our calculated canting angle of 1.6$^{\circ}$ for $U=5$ eV and $J=0$ eV severely underestimates the experimental value 
of 7.8$^\circ$\cite{jensen2009}.
In Fig.\ref{fig:LiMPO4} (a) we show the evolution of the canting angle with $J$ at $U=5$ eV.
We find that the canting angle is extremely sensitive to the value of $J$  -- in fact it is $\propto J^3$ -- changing from 
1.6$^\circ$ at $J=0$ eV to 7.8$^\circ$ at $J=1.7$ eV.
To reproduce the experimental value of the canting angle we need to use the rather large $J$ value of 1.7 eV.
The dependence of the canting angle on $J$ is consistent with Eqs.~\ref{Vupdn}, 
as the off-diagonal elements $n^{\uparrow\downarrow}$ and $n^{\downarrow\uparrow}$ are 
non-zero when the spins cant away from the easy axis.

In Fig.~\ref{fig:LiMPO4} (b) we report the energy versus the canting angle in LiNiPO$_4$ for $U=5$ eV and different values of $J$.
We see that as $J$ is increased from $J=0$ eV to $J=1$ eV (red circles and blue triangles) the minimum of the energy shifts to larger canting angle, with a stronger gain of energy with respect to the uncanted reference.
When performing the same calculation with $U_{eff}=$4 eV (green crosses in Fig.~\ref{fig:LiMPO4}) we obtain results that are very similar to the case $U=5$ eV and $J=0$ eV, which
is formally equivalent to the Dudarev approach with $U_{eff}=5$ eV. These comparisons confirm that varying $U$ has a minimal effect 
on the canting angle in LiNiPO$_4$ and also that the use of the Liechtenstein treatment of $J$ is extremly important.
To further confirm the direct relationship between the spin canting and the $J$ parameter, we performed the same calculations with $U=5$ eV and $J=1$ eV but we artificially fixed $J=0$ eV only in Eqs.~\ref{Vupdn}
(pink squares in Fig.\ref{fig:LiMPO4} (b)). We clearly see that the energy versus canting angle is strongly affected by this modification and in fact the canting is almost removed.

Similar $J$ dependence of the canting angle was also reported previously for 
Ni$^{2+}$ in BaNiF$_4$ \cite{ederer2006}; 
in Ref.~\onlinecite{ederer2006}
it was found that at $U=5$ eV, the canting varies from 2$^\circ$ to 3$^\circ$ 
when $J$ is varied from 0 eV to 1 eV.
In both LiNiPO$_4$ and BaNiF$_4$ the Ni ion 
is divalent, with a $d^8$ configuration, and octahedrally coordinated. 
To investigate the generality of this behavior, we next consider the case of the 
canted-spin antiferromagnet NiF$_2$, in which the Ni ion is in the same coordination environnement as in BaNiF$_4$.
Experimentally, NiF$_2$ has the spins aligned preferentially in the plane perpendicular to the \textit{c} axis with a slight canting from antiparallel 
alignment by an estimated $\sim$0.5$^\circ$ at low temperatures \cite{hutchings1970}.
Performing LSDA$+U$ calculations at the experimental volume and with $U=5$ eV and $J=0$ eV we indeed obtain the easy axis perpendicular to the \textit{c} axis and a small canting of 0.3$^\circ$, in excellent agreement with the experiments.
In contrast to the case of LiNiPO$_4$, however, we find that the amplitude of the canting angle is almost insensitive to the value of $J$ with just a small tendency to be reduced when $J$ increased.
This insensitivity of the canting angle to the value of $J$ in NiF$_2$ can be understood from the fact that in this compound the magnetism is almost collinear, and therefore the off-diagonal elements of the occupation matrix, $n^{\uparrow\downarrow}$ and $n^{\downarrow\uparrow}$, are close to zero. 
Inspection of Eqs.~\ref{Vupup} then shows that the effect of $J$ is reduced largely
to the diagonal part of the potential where the $U$ parameter is dominant.

To summarize our findings for the Ni-based compounds, in cases where the experimental
canting is large (2-3$^{\circ}$) we find a strong $J$-dependence of the canting 
angle, which increases with increasing $J$; when the canting is weak experimentally
the $J$-dependence is much weaker. 

\begin{widetext}
\begin{center}
\begin{figure}[ht]
\begin{minipage}[t]{1.0\textwidth}
 \includegraphics[width=4.4cm,keepaspectratio=true]{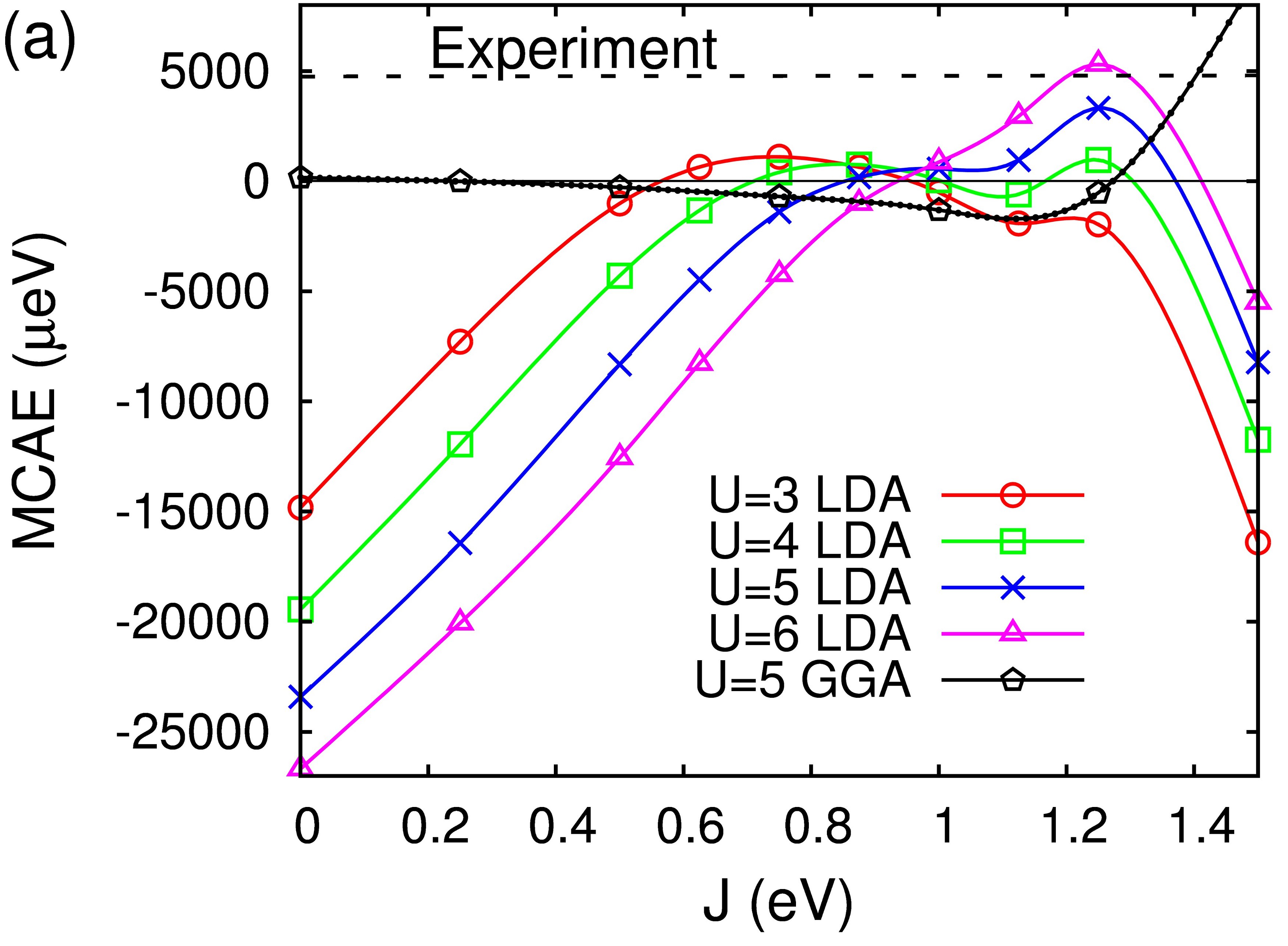}
 \includegraphics[width=4.4cm,keepaspectratio=true]{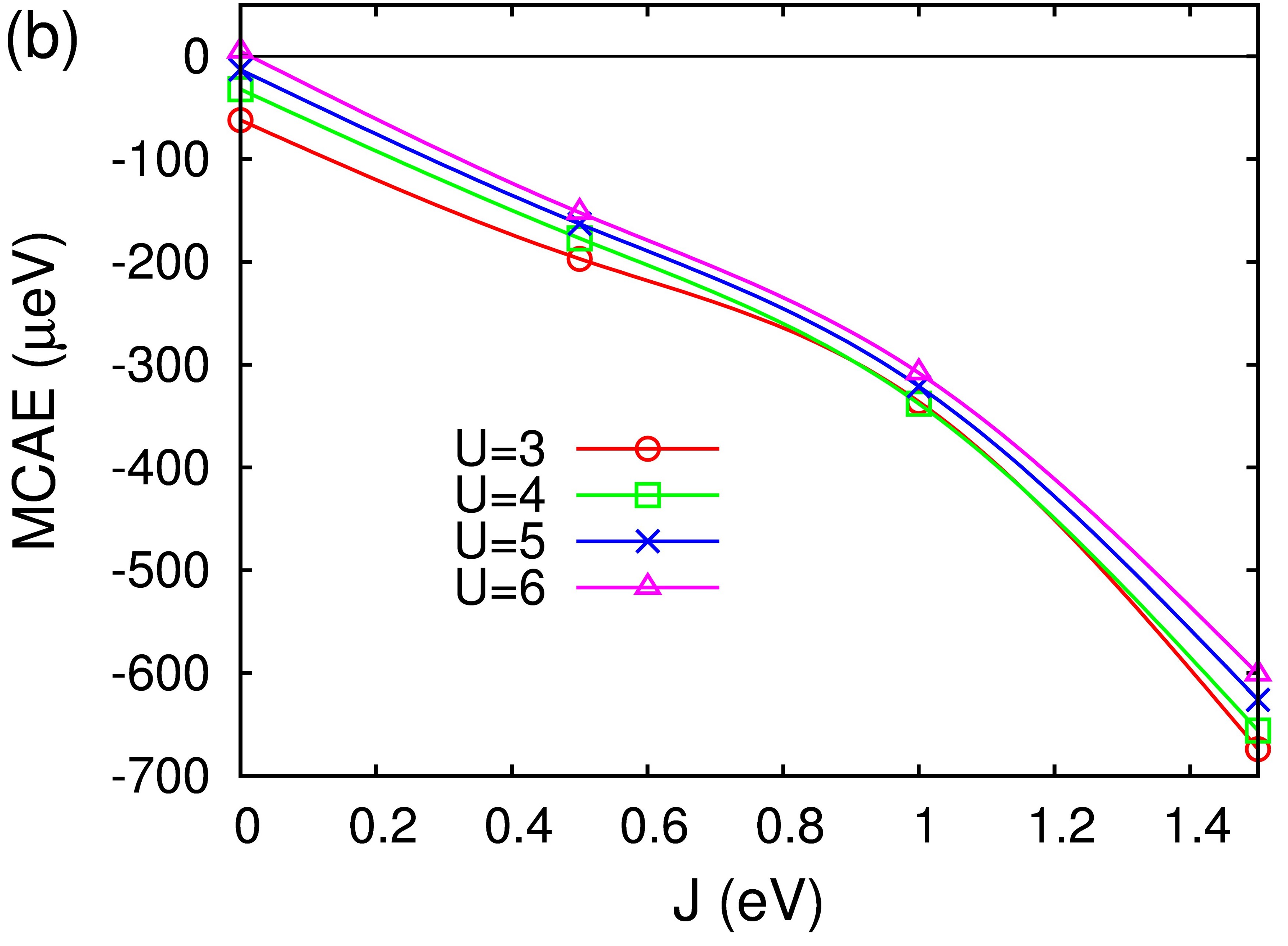}
 \includegraphics[width=4.4cm,keepaspectratio=true]{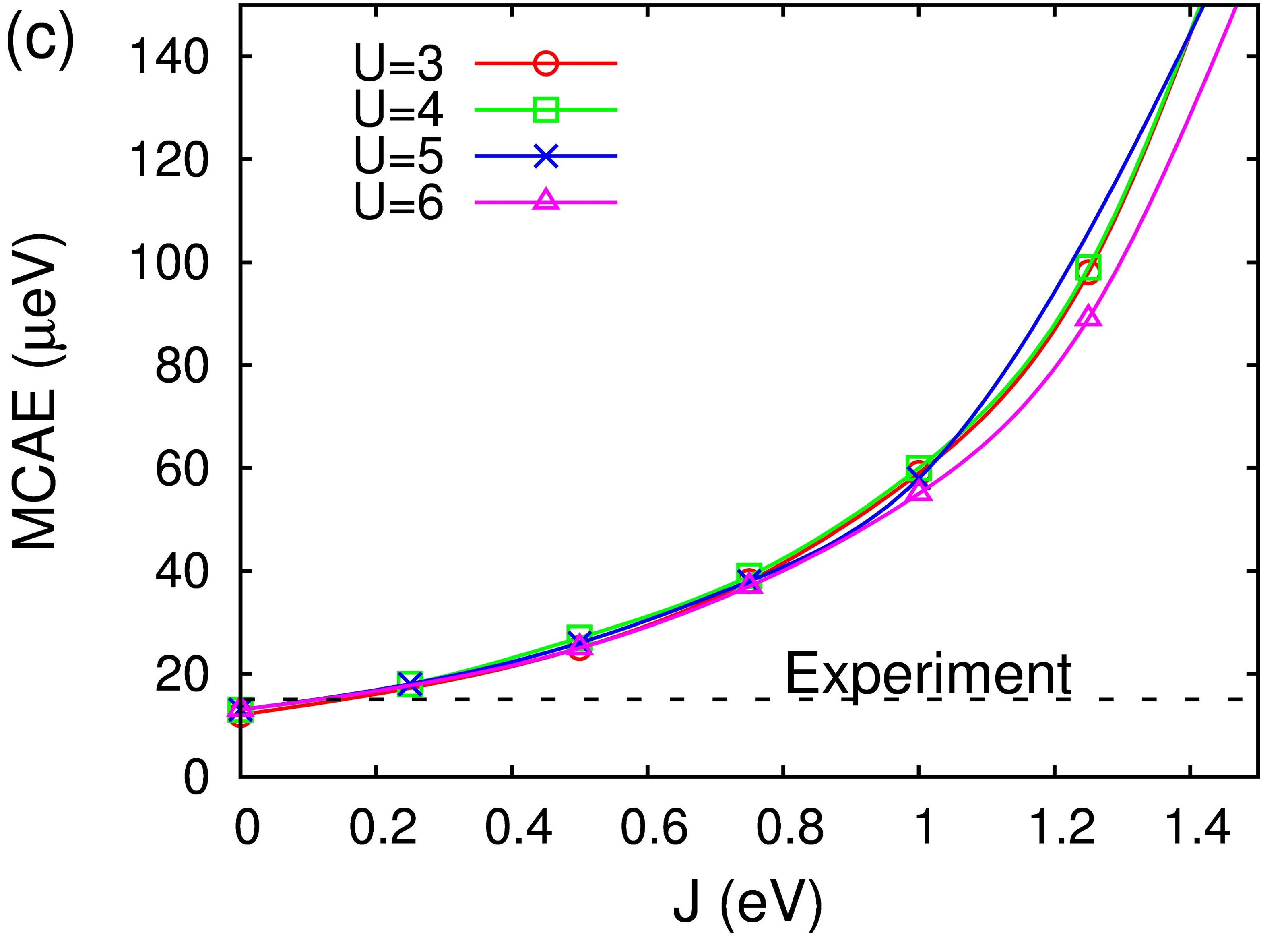}
 \includegraphics[width=4.4cm,keepaspectratio=true]{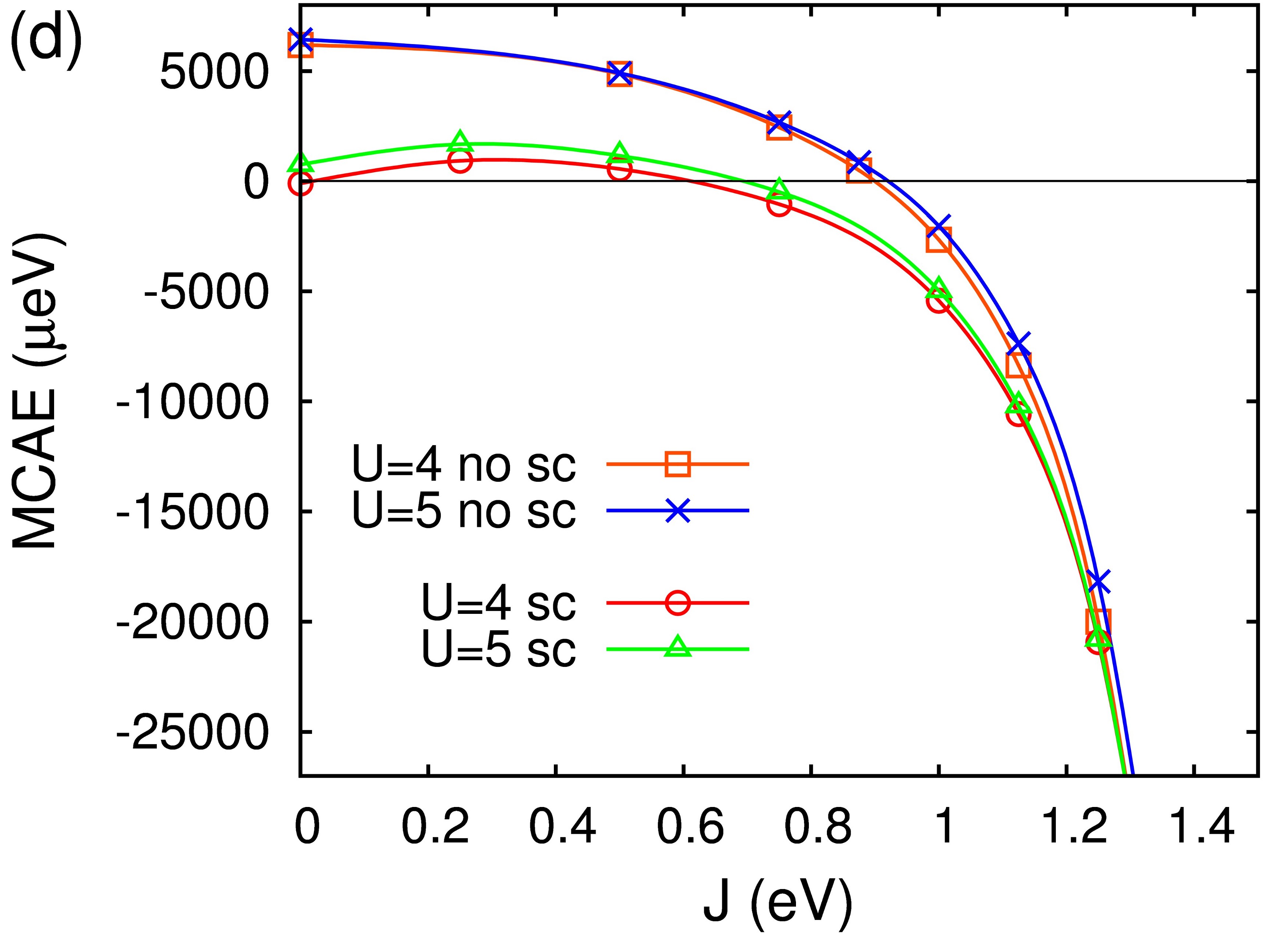}
 \caption{Magnetocrystaline anisotropy energy versus the $J$ parameter of (a) FeF$_2$ (Experimental value from Ref.\onlinecite{lines1967}), (b) NiF$_2$, (c) MnF$_2$ (Experimental value from Ref.\onlinecite{gafver1977}) and (d) CoF$_2$ (``sc`` are calculations with Co semi-cores while ''no sc'' are calculations without Co semi-cores).
The MCAE reported here is the energy between the $a$ and $c$ orientation of the spins, the energy of the $c$ orientation is taken as reference.}
\label{MCAE-MF2}
\end{minipage}
\end{figure}
\end{center}
\end{widetext}

Next we analyse the effect of $J$ on the behavior on the corresponding divalent iron compounds.
We begin with LiFePO$_4$, which is known
experimentally to be a $C$-type AFM with an easy axis along the $b$ direction and no 
observed canting of the spins~\cite{zimmermann2009,liang2008} ($C_y$ ground state with \textit{mmm'} magnetic point group).
Our calculations within the LSDA$+U$ functional at the commonly used values of $U=4$ eV and $J=0$ eV
for Fe$^{2+}$ yield the correct $C$-type AFM order but find the easy axis incorrectly
along the $a$ direction.
Now we switch to $J\neq0$ eV and report in Fig.~\ref{fig:LiMPO4}.c the MCAE
between the $b$ and $a$ directions, calculated by turning all the spins homogenously from the 
$C_y$ to the $C_x$ direction.
We find that the MCAE is approximately linear with $J$, but with
rather dramatic qualitative dependence: while at $J=0$ eV the easy axis is along the $a$ 
direction (negative MCAE) the MCAE is almost reduced to zero around 
$J=0.5$ eV and the easy axis changes to the $b$ direction for $J\gtrsim0.5$ eV (positive
MAE). To reproduce the experimental easy axis ($C_y$) a value of $J$ greater than 0.58 eV
is required. 
In the cases where the correct easy axis is reproduced ($C_y$) we do not observe any canting 
of the spins, in agreement with the experimental magnetic point group \textit{mmm'}.

As a second example with Fe$^{2+}$, we analyse the effect of $J$ on the MCAE of FeF$_2$.
Experimentally FeF$_2$ is known to have its spin magnetization parallel to the tetragonal $c$ axis with a rather large MCAE of about +4800 $\mu$eV~\cite{rudowicz1977,ohlmann1961}.
In Fig.~\ref{MCAE-MF2}.a we report the LSDA$+U$ MCAE energies with respect to $J$ at four different values of $U$ (3, 4, 5 and 6 eV).
All the calculations with $J=0$ eV give the wrong easy axis (spins are perpendicular to $c$) with a huge error in the MCA energy (MCAE from -16000 to -26000 $\mu$eV for $U$ going from 3 to 6 eV).
Increasing the value of $J$ in the range of 0--0.5 eV has the tendency to strongly reduce this error with a linear increase of the MCAE with $J$ as we found above for LiFePO$_4$.
However beyond $J\simeq0.5$ the increase of the MCAE is reduced and the evolution becomes more complex with the appearance of two maxima before a drastic decrease beyond $J\simeq1.3$ eV.
The correct easy axis (MCAE$>0$) is only obtained for a very small range of $U$ and $J$ values,
and the amplitude of the MCAE is correct over an even smaller range.
This $J$ dependence of the MCAE is again consistent with Eqs.\ref{Vupup}-\ref{Vupdn}.
From Eq.\ref{rho} it is clear that when changing the orientation of the spins from the $z$ axis to the $x$ or $y$ axis the off-diagonal parts of Eq.\ref{rho} become non-zero resulting in a direct effect of $J$ on the MCAE from Eqs.\ref{Vupdn}.

We also performed the same analysis of the MCAE for NiF$_2$ (Fig.\ref{MCAE-MF2}.b), MnF$_2$ (Fig.\ref{MCAE-MF2}.c) and CoF$_2$ (Fig.\ref{MCAE-MF2}.d).
MnF$_2$ and CoF$_2$ have the same easy axis as FeF$_2$ while NiF$_2$ has its easy axis perpendicular to the $c$ direction. 
The easy axis is well reproduced for all three compounds at $J=0$ eV.
As for FeF$_2$, the amplitudes of the MCAE depend strongly on $J$ but with a completely different trend in each compound.
For MnF$_2$ and FeF$_2$ the experimental value can be reproduced by adjusting the values of $U$ and $J$.
In the case of CoF$_2$ and NiF$_2$ no experimental values are available.
For CoF$_2$ we also performed calculations with and without Co semi-cores states (Fig.\ref{MCAE-MF2}.d) and find a strong difference in the magnitude of the MCAE for the two cases.
For FeF$_2$ we also performed calculations within the GGA functional (black pentagons in Fig.\ref{MCAE-MF2}.a) and obtained a completely different $J$ dependence than those calculated with the LDA functional.
These comparisons illustrate the difficulty of extracting a general rule about the $J$ dependence of the MCAE.

Our results reveal a problem with the predictability of the LSDA$+U$ method for
non-collinear magnetic materials: A strong dependence of the MCAE and spin canting angles on the values of $U$ and particularly $J$
that are used in the calculation.
Since properties such as magnetostriction, piezomagnetic response, magnetoelectric
response and exchange bias coupling are directly related to MCAEs and spin canting, 
it is of primary importance to reproduce these quantities accurately.
At the moment, the most reliable, although not entirely satisfactory, option 
appears to be a fine tuning 
of the $U$ and $J$ parameters by adjustment to reproduce experimentally measured
anisotropies and canting angles; there is some evidence to suggest that properties
such as magnetoelectric responses are then in turn well reproduced\cite{delaney2010}.
Future studies might explore methodologies for self-consistent calculation of the
$J$ parameter, or the predictions of new descriptions of the exchange and correlation
such as the hybrid functionals~\cite{heyd2004}. On the flip side, it is clear
that non-collinear magnetic systems provide a challenging case for testing the
correctness of new exchange correlation functionals within the density functional 
formalism. 

\textit{Acknowledgments:}
% We are grateful for fruitful dis-cussions with B. Amadon. 
This work was supported by the Department of Energy SciDAC DE-FC02-06ER25794.
We made use of computing facilities of TeraGrid at the National Cen-ter for Supercomputer Applications and of the California
Nanosystems Institute with facilities provided by NSF grant No. CHE-0321368 and Hewlett-Packard.
EB also acknowledges FRS-FNRS Belgium and the ULg SEGI supercomputer facilities.

% \bibliography{biblio}
 
%Merlin.mbs v4.21 2009-07-09.
%

\end{document}